\documentclass[preprint,showpacs,preprintnumbers,amsmath,amssymb]{revtex4}
\usepackage[english]{babel}
\usepackage{bm} 
\usepackage{graphicx}

\newcommand{\beq}{\begin{equation}}
\newcommand{\eeq}{\end{equation}}
\newcommand{\bea}{\begin{eqnarray}}
\newcommand{\eea}{\end{eqnarray}}

\begin{document}
\selectlanguage{english}

\title{On the Origin of the $\log d$ Variation of the Electrostatic Force Minimizing Voltage in Casimir Experiments}

\author{\firstname{S.~K.}~\surname{Lamoreaux}}
\email{steve.lamoreaux@yale.edu}
 \affiliation{Yale University, Department of Physics, P.O. Box 208120, New Haven, CT 06520-8120}
\author{\firstname{A.~O.}~\surname{Sushkov}}
\email{alex.sushkov@yale.edu}
 \affiliation{Yale University, Department of Physics, P.O. Box 208120, New Haven, CT 06520-8120}

\begin{abstract}
A number of experimental measurements of the Casimir force have observed a logarithmic distance variation of the voltage that minimizes electrostatic force between the plates in a sphere-plane geometry.  We show that this variation can be simply understood from a geometric averaging of surface potential patches together with the Proximity Force Approximation.
\end{abstract}

\maketitle

A number of experimental measurements of the Casimir force have observed a distance variation of the voltage applied between the plates that minimizes the electrostatic potential.\cite{kim1,kim2,deman,deman2}  This distance variation is of the approximate form 
\begin{equation}
V_m(d)=a+b\log d
\end{equation}
over a range of distances $d$ spanning up to nearly two orders of magnitude.
We have shown numerically that a variation in the minimizing voltage can result from the geometrical averaging of patch potentials on the plate surfaces, and have developed a heuristic explanation of the effect, as described in \cite{kim}.  However the $\log d$ form of the variation was not explained or derived in \cite{kim}.  Here we provide added details showing that in the case of a spherical surface/plane surface geometry, the $\log d$ variation arises quite naturally.

The electrostatic force between the plates is minimized when the free energy is minimized.  For a plate with a spherical surface curvature $R$ together with a flat plate, both of diameter $2R_m$, the Proximity Force Approximation (PFA)  works well for distances $d< R_m^2/2R$; in addition, the characteristic radius of the patches $r_0$ must satisfy $r_0>\sqrt{2Rd}$, in which case we can consider each patch as only interacting with its own image in the opposite plate.  If this latter criterion is not met, the effects calculated here will be reduced in magnitude, however the general conclusions are otherwise unaltered.

For random patches on the surfaces, the electrostatic free energy is
\begin{equation}
U=\sum {1\over 2} CV^2={\epsilon_0 R\over 2} \int_0^{R_m}\int_0^{2\pi} (V_a-V_p(r,\phi))^2{(r/R)\ d\phi\  dr\over d+r^2/2R}
\end{equation}
where $V_a$ is a voltage applied between the plates, and $V_p(r,\phi)$ describes the random voltage patches on the plates' surfaces.  The value of $V_a$ that minimizes the force (or electrostatic free energy) is referred to as the minimizing potential, and is often called the contact potential.  $V_m$ is found by taking the derivative with respect to $V_a$:
\begin{equation}\label{vm}
{\partial U\over \partial V_a}=0={\epsilon_0 R}\int_0^{R_m}\int_0^{2\pi} (V_a-V_p(r,\phi)){(r/R)\ d\phi\  dr\over d+r^2/2R}\ \ {\rm for\ } V_a=V_m.   
 \end{equation}
Because the only $\phi$ dependence is in $V_p(r,\phi)$, the angular integral of this term can be replaced with $V_p(r)$, with the  understanding that this replacement represents a sum of the individual patches, of typical radius $r_0$, intersected by a circle of radius $r$.  For homogeneous random patches, this number is roughly $(2\pi r)/(2 r_0)$, for $r>r_0$, or 1 for $r<r_0$. Then if $RMS$ magnitude of the patches is $V_0$ with zero average, this sum will have magnitude proportional to the square root of the number of patches in the sum times $\pm V_0$, or 
\begin{equation}
\int_0^{2\pi} (V_m-V_p(r,\phi))d\phi =2\pi \left[V_m+V_p(r)\right]
\end{equation}
\begin{equation}
\approx 2\pi  V_m\pm  \left[\Delta\phi\left[\theta(r-r_0)-\theta(r-R_m)\right]\sqrt{2\pi r\over 2 r_0}+2\pi\left[\theta(r)-\theta(r-r_0)\right]\right] V_0 
\end{equation}
where $\theta(r-r_0)$ is the Heaviside step function, $\Delta\phi=2\pi r_0/r$ results from $d\phi$ in the integral, and a single patch near the center of the plates is included. (The relative sign of these two terms on the r.h.s. is random.) Thus, $V_p(r)\propto 1/\sqrt{r/r_0}$ for $r\gg r_0$.  Although we will not directly use this result in the subsequent discussion, it is important to note that $|V_p(r)|\leq |V_p(0)|$ in the case of homogenous random patches.  

Equation (\ref{vm}) can be integrated by parts, yielding 
\begin{equation}
(V_m-V_p(r))\log(d+r^2/2R)|_0^{R_m} +\int_0^{R_m}\log(d+r^2/2R){dV_p(r)\over d r} dr=
\end{equation}
\begin{equation}
=(V_m-V_p(R_m))\log(d+R_m^2/2R)-(V_m-V_p(0))\log d -Q(d)=0
\end{equation}
where we have introduced a new function $Q(d)$. Then 
\begin{equation}
V_m(d)={V_p(R_m)\log(d+R_m^2/2R)-V_p(0)\log d +Q(d)\over \log(d+R_m^2/2R)-\log d}.
\end{equation}

There are three cases to consider.  

\noindent
{\bf Case 1: Close range:} $d<< R_m^2/2R$, $|\log d|>> |\log R_m^2/2R|$ and $d<<r_0^2/2R$

In this limit, one single patch at or near the center dominates in the determination of $V_m$, and $dV_p(r)/dr\approx 0$ when $r^2/2R<d$ in the integral defining $Q(d)$.  The $\log d$ terms have the largest magnitude, and thus we can neglect all terms not multiplied by $\log d$.  In this case,
\begin{equation}
V_m(d)=V_p(0).
\end{equation}

\noindent
{\bf Case 2: Intermediate range:} $d<< R_m^2/2R$, $|\log d|< |\log R_m^2/2R|$ and $d<r_0^2/2R$

In this case,
\begin{equation}
\log(d+R_m^2/2r)\approx \log(R_m^2/2R)
\end{equation}
so
\begin{equation}
Q(d)=Q_0\log(R_m^2/2R)
\end{equation}
because in this limit $Q(d)$ is nearly independent of $d$.  This can be seen in Eq. (5) where the $\log(d+r^2/2R)$ factor contributes a $d$ dependence only when $r$ is small, but in that limit, $dV_p(r)/dr\approx 0$. 
Therefore, 
\begin{equation}
V_m(d)\approx {V_p(R_m)\log(R_m^2/2R)-V_p(0)\log d +Q_0 \log(R_m^2/2R) \over \log(R_m^2/2R)-\log d}.
\end{equation}
The denominator can be Taylor expanded and we therefore arrive at
\begin{equation}
V_m(d)\approx \left[V_p(R_m)-{V_p(0)\log d\over \log(R_m^2/2R)} + Q_0\right]\left[1+{\log d\over \log(R_m^2/2R)}\right]\approx a+b\log d
\end{equation}
where $a$ and $b$ do not depend on $d$, and where terms only first order in $\log d/\log(R_m^2/2R)$ are retained.  In addition, there can be a contribution to $a$ from external circuit contact potentials.

It should be noted that a single patch at large $r$ will generate a non-zero $Q_0$. Depending on the size and magnitude of the single patch, it might dominate the $\log d$ distance dependence of $V_m$.

\noindent
{\bf Case 3:} $d> R_m^2/2R$ and $d\ {\buildrel > \over \sim}\  
r_0^2/2R$ (so that the PFA remains valid)

For this case, it is easiest to go back to Eq. (3) and expand the denominator.  We have
\begin{equation}
0=\int_0^{R_m}\int_0^{2\pi}(V_m-V_p(r,\phi))r(1-r^2/2Rd)drd\phi\approx \int_0^{R_m}\int_0^{2\pi}(V_m-V_p(r,\phi)) r dr d\phi
\end{equation}
so that
\begin{equation}
V_m={\int_0^{R_m}\int_0^{2\pi} V_p(r,\phi) r dr d\phi\over \pi R_m^2}=\langle V_p(r,\phi)\rangle
\end{equation}
which is the surface average of $V_p(r,\phi)$.

\noindent
{\bf Discussion}

It is easy to understand the first and third cases where $d\rightarrow 0$ and $d\rightarrow \infty$. In the first case, a single patch dominates the electrostatic force, and its potential determines $V_m$. In the latter, $V_m$ is simply the average surface potential, as the variation is the distance between the surfaces due to the curvature is very small compared to $d$.

The intermediate case is slightly more difficult to understand, but arises from, with increasing distance, the loss in dominance (in magnitude) of $\log d$, compared to $\log (d+R_m^2/2R)$.  

Numerical calculations based on Eq. (3), using random patches specified on a surface, are straightforward and fully support the essential conclusions presented above.  As an aside, there is a remaining question as to where it is the energy or force that must be minimized.  Numerical calculations based on the minimization of the force produces no statistically significant differences compared to the energy minimization, as might be expected.

Of course, this is a very simplistic model, particularly in the assumptions that the patch potentials are randomly $\pm V_0$, that all patches have the same radius and are circular, and that the radius of the positive patches is equal to that of the negative patches.  These assumptions are not essential to the derivation of Eq. (8), which is the principal result reported here. Further refinements will likely not significantly change the subsequent conclusions presented here.  For example, as discussed already, a single patch at large $r$ will generate a non-zero $Q_0$. If this patch has a large area and/or large potential ({\it e.g.}, a charged speck of dust or a scratch), it could easily be the dominant contribution to the $\log d$ distance dependence.  For such a patch, $V_m(d)$ will only vary slowly with a relative translational repositioning of the plates.

This work was supported by the DARPA/MTOs Casimir Effect Enhancement project under SPAWAR Contract No. N66001-09-1-2071.



\begin{thebibliography}{99}
 
\bibitem{kim1} W.-J. Kim, M. Brown-Hayes, D.A.R. Dalvit, J.H. Brownell, and R. Onofrio, Phys. Rev. A {\bf 78}, 020101(R) (2008).
\bibitem{kim2} W.-J. Kim, A.O. Sushkov, D.A.R. Dalvit, and S.K. Lamoreaux, Phys. Rev. Lett.  {\bf 103}, 060401 (2009).
\bibitem{deman} S. de Man, K. Heeck, and D. Iannuzzi, Phys. Rev. A {\bf 79}, 024102 (2009).
\bibitem{deman2} S. de Man, K. Heeck, R.J. Wijngaarden, and D. Iannuzzi, J. Vac. Sci. Tech. B {\bf 28}, C4A25 (2010).
\bibitem{kim} W.J. Kim, A.O. Sushkov, D.A.R. Dalvit, and S.K. Lamoreaux, Phys. Rev. A {\bf 81}, 022505 (2010).

\end{thebibliography}
\end{document}